\documentclass[twocolumn, tighten, times]{aastex61}
\usepackage{graphicx}
\usepackage{epstopdf}

\usepackage{color}
\usepackage{soul}

\renewcommand\baselinestretch{1.3}
\shorttitle{The morphological quenching mechanism}
\shortauthors{Lu \& Fang et al.}
\begin{document}

\title{The effect of the morphological quenching mechanism on star formation activity \\ at $0.5 < z < 1.5$ in 3D-HST/CANDELS}

\correspondingauthor{Guanwen Fang}
\email{wen@mail.ustc.edu.cn, xkong@ustc.edu.cn}

\author{Shiying Lu}
\affil{School of Mathematics and Physics, Anqing Normal University, Anqing 246011, People's Republic of China}
\affil{School of Astronomy and Space Science, Nanjing University, Nanjing 210093, People's Republic of China}

\author{Guanwen Fang}
\altaffiliation{Guanwen Fang and Shiying Lu contributed equally to this work.}
\affil{School of Mathematics and Physics, Anqing Normal University, Anqing 246011, People's Republic of China}

\author{Yizhou Gu}
\affil{School of Physics and Astronomy, Shanghai Jiao Tong University, Shanghai 200240, People’s Republic of China}

\author{Qirong Yuan}
\affil{School of Physics Science and Technology, Nanjing Normal University, Nanjing  210023, People’s Republic of China}

\author{Zhen-Yi Cai}
\affil{CAS Key Laboratory for Research in Galaxies and Cosmology, Department of Astronomy, University of Science and Technology of China, Hefei, 230026, People's Republic of China}
\affil{School of Astronomy and Space Science, University of Science and Technology of China, Hefei 230026, People’s Republic of China}

\author{Xu Kong}
\affil{CAS Key Laboratory for Research in Galaxies and Cosmology, Department of Astronomy, University of Science and Technology of China, Hefei, 230026, People's Republic of China}
\affil{School of Astronomy and Space Science, University of Science and Technology of China, Hefei 230026, People’s Republic of China}

\begin{abstract}
Several mechanisms for the transformation of blue star-forming to red quiescent galaxies have been proposed, and the green valley (GV) galaxies amid them are widely accepted in a transitional phase. Thus, comparing the morphological and environmental differences of the GV galaxies with early-type disks (ETDs; bulge dominated and having a disk) and late-type disks (LTDs; disk dominated) is suitable for distinguishing the corresponding quenching mechanisms. A large population of massive ($M_* \geqslant 10^{10}M_\odot$) GV galaxies at $0.5 \leqslant z \leqslant 1.5$ in 3D-HST/CANDELS is selected using extinction-corrected $(U-V)_{\rm rest}$ color.
After eliminating any possible active galactic nucleus candidates and considering the ``mass-matching'', we finally construct two comparable samples of GV galaxies with either 319 ETD or 319 LTD galaxies.
Compared to the LTD galaxies, it is found that the ETD galaxies possess higher concentration index and lower specific star formation rate, whereas the environments surrounding them are not different. This may suggest that the morphological quenching may dominate the star formation activity of massive GV galaxies rather than the environmental quenching. To quantify the correlation between the galaxy morphology and the star formation activity, we define a dimensionless morphology quenching efficiency $Q_{\rm mor}$ and find that $Q_{\rm mor}$ is not sensitive to the stellar mass and redshift. When the difference between the average star formation rate of ETD and LTD galaxies is about 0.7 $M_\odot \rm \;yr^{-1}$,  the probability of $Q_{\rm mor}\gtrsim 0.2$ is higher than 90\%, 
which implies that the degree of morphological quenching in GV galaxies might be described by $Q_{\rm mor}\gtrsim 0.2$. 
\end{abstract}


\section{Introduction}
\label{sec1:introduction}
With the development of the high-quality all-sky galaxy surveys, a conspicuous bimodality has been discovered on various diagrams, such as on the color-color diagram (e.g., \citealt{Strateva+01, Hogg+02, Blanton+03a}), on the color-magnitude (e.g., \citealt{Graves+09}) or color-mass diagram (e.g., \citealt{Schawinski+14}), and on the star formation rate (SFR)-mass diagram (e.g., \citealt{Schiminovich+07}). Two distinct galaxy populations with red and blue colors have been commonly referred to as the red sequence (RS) and the blue cloud (BC), respectively. The intermediate zone between the two main populations, commonly known as the green valley (GV), has been viewed as the crossroads of galaxy evolution (e.g., \citealt{Salim+07,Gu+18}).

By the description in a series of papers (e.g., \citealt{Faber+07,Martin+07, Schawinski+07,Thomas+10,Schawinski+14,Pandya+17,Gu+18}), the galaxy may cross over the GV in many possible different evolutionary pathways. On one hand, the GV galaxies can be undergoing the quenching process of star formation, which moves a galaxy from the BC to the RS. To explore the evolutionary connection among red, green, and blue galaxy populations, \cite{Gu+18} investigate different physical properties of these three populations at $0.5<z<2.5$ in the five 3D-HST/CANDELS fields, including dust content, morphologies, structures, active galactic nucleus (AGN) fraction, and environments. Their findings favor the scenario that the GV galaxies are at a transitional phase from the BC to the RS, which is consistent with multiple survey-based studies (e.g., \citealt{Schiminovich+07,Pandya+17}).
On the other hand, the GV galaxies may be caused by rejuvenation events from infalling of gas clouds or gas-rich smaller galaxies. \cite{Thomas+10} and \cite{Chauke+19} find that rejuvenating galaxies have undergone some secondary star-formation events at low and high redshifts. The rejuvenation event can also give a possible explanation for some GV galaxies showing blue outer regions (\citealt{Thilker+10,Salim+Rich+10,Fang+12}).

The time-scale of a galaxy across the GV relies on the evolutionary process of the galaxy itself, accompanying morphological transformation (e.g., \citealt{Wuyts+11b}). Utilizing galaxies at $0.1<z<0.2$ from the Galaxy and Mass Assembly (GAMA) survey,  \cite{Phillipps+19} derive that the evolutionary time-scales of typical low-redshift galaxies from the BC through the GV to the RS are about 2-4 Gyr. In contrast, \cite{Pandya+17} place an observational upper limit on the average population transition time-scale as a function of redshift, finding that galaxy quenching is on a fast track at high redshift, whereas it is on a slow track at low redshift, which is also consistent with the results from \cite{Gu+19}. Note that the quenching should be correlated with the morphology of the galaxy, if the quenching mechanism involves the reorganization of the stellar component of the galaxy via compaction events or dissipative mergers (\citealt{Dekel+09,Barro+13,Barro+17, Wellons+15}) or a natural consequence of `inside-out' growth combined with disk fading (e.g., \citealt{Lilly+Carollo+16}). \cite{Schawinski+14} analyze the morphology of GV galaxies, and reveal two evolutionary pathways toward quenching in early-type galaxies (ETGs) and late-type galaxies (LTGs): a short quenching timescale ($\sim$100 Myr) for ETGs  and a longer timescale ($\sim$2-3 Gyr) for LTGs.

Contrary to the common quenching mechanisms, used to remove the cold gas reservoir or shut down any fresh supply of cold gas, such as merger or disk instabilities (e.g., \citealt{Toomre+Toomre+1972, Dekel+09}) and some feedback (e.g., \citealt{Cai+13, Brennan+17}), ``morphological quenching'' (MQ), proposed by \cite{Martig+09}, has usually been regarded as a quenching mechanism of star formation, where once the spheroid has been built up via early minor merger the gas disk
can be stabilized against star formation without an external mechanism to remove its present gas or terminate its fresh gas supply. The gravitational potential of a centrally concentrated bulge not only can help stabilize the gas against fragmenting, suggested by simulations from \cite{Su+19} and \cite{Gensior+20}, but also can decrease the strength of bars and the efficiency of gas consumption toward the galactic center (\citealt{Barazza+08, Fragkoudi+16}), finally leading to a reduced SFR (\citealt{Zurita+04,Sheth+05}).
The natural process of MQ proposes a simple explanation for the local observations of some red ETGs hosting significant amounts of atomic or molecular gas with high densities (e.g., \citealt{Morganti+06, Crocker+09, Young+09b, Donovan+09}), and the conspicuously observed bimodality of galaxies.

The GV galaxy is thus a suitable sample to be studied because it can make sense to compare the morphological transition of ETGs and LTGs in the GV at the same evolutionary stage.
Recently, on the basis of the CO($J$ =1-0) observation of nearby ($z<0.05$) GV galaxies, \cite{Koyama+19} assess the effects of GV galaxy morphologies (i.e., disk- and bulge-dominated GV galaxies) on the efficiency of star formation, and find that the emergence of the stellar bulge does not decrease the efficiency of ongoing star formation, which contrasts with the prediction of the MQ (e.g., \citealt{Martig+09, Bluck+14}).
But conceivably, the MQ could be even more important at high redshift, which results from the fact that the RS is already in place (e.g., \citealt{Gu+18,Gu+19}), and the gas fraction in ETGs might be higher owing to the higher gas infall rates and the greater presence of unvirialized clusters (e.g., \citealt{Bell+04,Cirasuolo+07}). Therefore, it is still desirable to assess the effects of galaxy morphologies on the star formation quenching at higher redshift.

In order to discern the real impact of the MQ mechanism on star formation activity, especially at high redshift, we utilize the extinction-corrected rest-frame $U$-$V$ color (\citealt{WangT+17}) to construct a sample of massive ($M_* \geqslant 10^{10}M_\odot$) GV galaxies at $0.5\leqslant z\leqslant 1.5$ in the five fields of 3D-HST/CANDELS (\citealt{Grogin+11, Koekemoer+11, Skelton+14}). After excluding the influence of AGNs as much as possible, the sample of GV galaxies is further separated into ETGs and LTGs, based on a catalog of visual-like H-band morphologies estimated using convolutional neural networks, which is provided by \cite{Huertas-Company+15b}.
We then compare the concentration, the specific SFR (sSFR), the morphology quenching efficiency ($Q_{\rm mor}$), and the  surrounding environment of ETGs and LTGs to gradually clarify the effect of MQ on the star formation activity.

The layout of this paper is as follows.
The data set and our sample construction are described in Section \ref{sec2:data and sample}. We analyze the MQ on star formation activity for early-type disk (ETD) and late-type disk (LTD) galaxies in GV galaxies in Section \ref{sec3:analysis} through comparisons of the concentration in Section \ref{sec3.1:Conc}, the sSFR in Section \ref{sec3.2:sSFR}, the MQ efficiency in Section \ref{sec3.3:Qmor}, and the environment in Section \ref{sec3.4:env}. Discussion on the MQ is presented in Section \ref{sec4:discussion}. The final summary is given in Section \ref{sec5:summary}.
Throughout our paper, we adopt the cosmological parameters as follows: $H_0=70\,{\rm km~s}^{-1}\,{\rm Mpc}^{-1}$, $\rm \Omega_m=0.30$, and $\Omega_{\Lambda}=0.70$.

\section{Data and Sample}
\label{sec2:data and sample}
\subsection{3D-HST/CANDELS}
\label{sec2.1:3d-hst/candels}
The 3D-HST and CANDELS Multi-Cycle Treasury programs have provided WFC3 and ACS spectroscopy and photometry in five fields in a homogenous way, including the AEGIS, COSMOS, GOODS-North, GOODS-South, and UKIDSS UDS fields (\citealt{Grogin+11, Koekemoer+11, Skelton+14}). The coverage is over 900 arcmin$^2$ which is helpful to relieve the effect of cosmic variance. A wealth of the homogeneous imaging data benefits construction of the spectral energy distributions (SEDs) of objects over a wide wavelength range.

The corresponding derived data products are provided by \cite{Skelton+14}, including the photometric redshift, the rest-frame colors, and the stellar population parameters. \cite{Skelton+14} use the EAZY code (\citealt{Brammer+08}) to determine the photometric redshift by fitting the SED of each galaxy with a linear combination of seven galaxy templates. In this work, we preferentially take the spectroscopic redshift ($z_{\rm spec}$) if available; otherwise, we use the photometric redshift ($z_{\rm phot}$). The $z_{\rm phot}$ estimates have a high precision, with the largest (lowest) $\sigma_{\rm NMAD}$=0.026 (0.007) for GOODS-N (COSMOS). Additionally, \cite{Skelton+14} utilize the FAST code (\citealt{Kriek+09}) to derive stellar population parameters, such as the stellar mass ($M_*$) and dust attenuation ($A_{\rm V}$), adopting the \cite{Bruzual+Charlot+03} stellar population synthesis (SPS) models with a \cite{Chabrier+03} initial mass function (IMF) and solar metallicity. While it is more reasonable for high-redshift star-forming galaxies (SFGs) to take the contribution of the asymptotic giant branch (AGB) stars into consideration, \cite{WangT+17} thus adopt the \cite{Maraston+05} SPS models instead of \cite{Bruzual+Charlot+03} SPS models to better estimate the dust attenuation ($A_{\rm V}$). In order to ensure the consistency on the definition of the GV, we also perform the FAST code to reassess the stellar population parameters, following \cite{WangT+17}. This means that, assuming an exponentially declining star-forming history and the \cite{Calzetti+00} extinction, the \cite{Maraston+05} SPS models with a \cite{Kroupa+01} IMF and solar metallicity are taken instead. As described in \cite{Gu+18}, comparing the values of stellar mass and $A_{\rm V}$ derived with the \cite{Bruzual+Charlot+03} SPS models and the \cite{Chabrier+03} IMF to those derived with \cite{Maraston+05} SPS models and the \cite{Kroupa+01} IMF, the difference is small. The average values of stellar mass ($A_{\rm V}$) for the latter are 0.05 dex (0.18 dex) lower than those for the former, which is due to the contribution of AGB stars.

The SFR of a galaxy is estimated by combing contributions of the ultraviolet (UV) light from massive stars and of the infrared (IR) emissions of dust-reprocessed light from an older stellar population. Both UV and IR emissions have been provided by \cite{Whitaker+14}. Assuming a \cite{Chabrier+03} IMF and the \cite{Bell+05} UV conversion, the total SFR can be derived by
\begin{equation}\label{eq1-SFR}
{\rm SFR_{UV+IR}}[M_\odot \cdot {\rm yr^{-1}}]=1.09\times10^{-10}
				\times (L_{\rm IR}+2.2 L_{\rm UV})/ L_{\odot},
\end{equation}
where $L_{\rm UV}$ represents the integrated rest-frame UV (1216-3000~\AA) luminosity, defined by a factor of 1.5 times the 2800~\AA~ rest-frame monochromatic luminosity, $L_{\rm UV} = 1.5 \times L_{\rm 2800}$. The IR luminosity, $L_{\rm IR}$, represents the integrated 8-1000~$\rm \mu m$ luminosity, which is converted from the Spitzer/MIPS 24~$\rm \mu m$ data using a single luminosity-independent template \citep{Dale+Helou+02,Wuyts+08,Franx+08}.
If the 24~$\rm \mu m$ detection is unavailable, the SFR will be corrected by dust attenuation ($A_{\rm v}$), assuming the \citet{Calzetti+00} dust attenuation curve:
\begin{equation}\label{eq2-SFR}
  {\rm SFR_{UV,corr}} [M_{\odot} \cdot {\rm yr^{-1}}] = {\rm SFR_{UV}} \times 10^{0.4 \times 1.8 \times A_{\rm v}},
\end{equation}
where ${\rm SFR_{UV}}= 3.6 \times 10^{-10} \times L_{\rm 2800}/ L_{\odot}$, assuming the \citet{Chabrier+03} IMF, which is following \citet{Wuyts+11a}.
The ${A_{\rm v}}$ is derived during the SED fitting with the FAST code. The acceptable agreement between ${\rm SFR_{UV,corr}}$ and ${\rm SFR_{UV+IR}}$  is confirmed by previous works (e.g., \citealt{Rangel+14,Fang+18,Gu+19,Gu+20}).

\subsection{Morphology}
\label{sec2.2:morphology}
The kinematics and evolutionary history of a galaxy are all relevant to its morphology \citep{Kormendy+Kennicutt+04}.
\cite{Huertas-Company+15b} present a catalog of visual-like $H$-band morphologies of $\sim$50,000 galaxies in the five CANDELS fields. Morphologies are estimated by mimicking human perception with deep learning using convolutional neural networks.
For each galaxy there are five parameters $f_{\rm spheroid}$, $f_{\rm disk}$, $f_{\rm irr}$, $f_{\rm PS}$, and $f_{\rm Unc}$, which refer to the probabilities of having a spheroid, a disk, some irregularities, being a point source (or unresolved), and being unclassifiable, respectively. From the morphological analysis of \cite{Gu+18} (see their Section 4 for details), in the same five 3D-HST/CANDELS fields, it can be found that there are more than 50$\%$ of GV galaxies at $0.5\leqslant z<1.5$ that classified as the ETD and LTD types. Therefore, following the classification of \cite{Huertas-Company+15b}, we utilize the thresholds in the different probabilities to select the ETD galaxies (bulge dominated and having a disk) and LTD galaxies (disk dominated) at $0.5\leqslant z\leqslant 1.5$, for the purpose of a more comprehensive analysis on the MQ in the GV galaxies. The definitions of classification for ETD and LTD galaxies are presented here:\\
\\
1. ETDs: $f_{\rm spheroid}>2/3$, $f_{\rm disk}>2/3$, and $f_{\rm irr}<0.1$;\\
2. LTDs:  $f_{\rm spheroid}<2/3$, $f_{\rm disk}>2/3$, and $f_{\rm irr}<0.1$.\\

To quantitatively describe the bulge of a galaxy, we perform the nonparametric measurements on the near-IR (NIR) images, using the \textit{Morpheus} software developed by \cite{Abraham+07}. The nonparametric parameter refers to the concentration index ($C$), which can describe the concentration of the surface brightness distribution of a galaxy. Many previous works have tested and adopted this measurement (\citealt{Kong+09,Wang+12,Fang+15,Gu+18,Gu+19,Lu+20}). The nonparametric measurements are sensitive to the signal-to-noise ratio ($\rm S/N$) per pixel, especially for $\rm S/N \leqslant 2$ (\citealt{Lotz+04,Lisker+08}), while one of our selection criteria, $\tt use\_phot$=1, can ensure that our measurements are performed on a reliable detection in the $H_{\rm F160W}$ (see section \ref{sec2.3:sample selection}). Therefore, the measurement of the concentration index, $C$, is reliable and should not suffer from the signal-to-noise ratio ($\rm S/N$) effect.

\subsection{Sample Selection}
\label{sec2.3:sample selection}
We make use of a large set of multiwavelength photometric data from 3D-HST/CANDELS to construct a sample of massive ($M_* \geqslant 10^{10}M_\odot$) galaxies with reliable detection ($\tt use\_phot$=1) at $0.5\leqslant z \leqslant1.5$. The flag of $\tt use\_phot$=1 can ensure that the source is a galaxy and not close to a bright star and is detected in at least two individual exposures in each of the F125W and F160W bands. Meanwhile, the flag also means that the detected source has an $\rm S/N>3$ in the F160W with ``noncatastrophic'' fits on both the photometric redshift and stellar population properties (see section 3.8 in \citealt{Skelton+14}). The completeness above the mass threshold, $M_* \geqslant 10^{10}M_\odot$, is $\sim$90\% up to the highest redshift considered (\citealt{Grogin+11,Wuyts+11a,Newman+12,Barro+13,Pandya+17}).

\begin{figure*}[htb]
\center{\includegraphics[width=15cm] {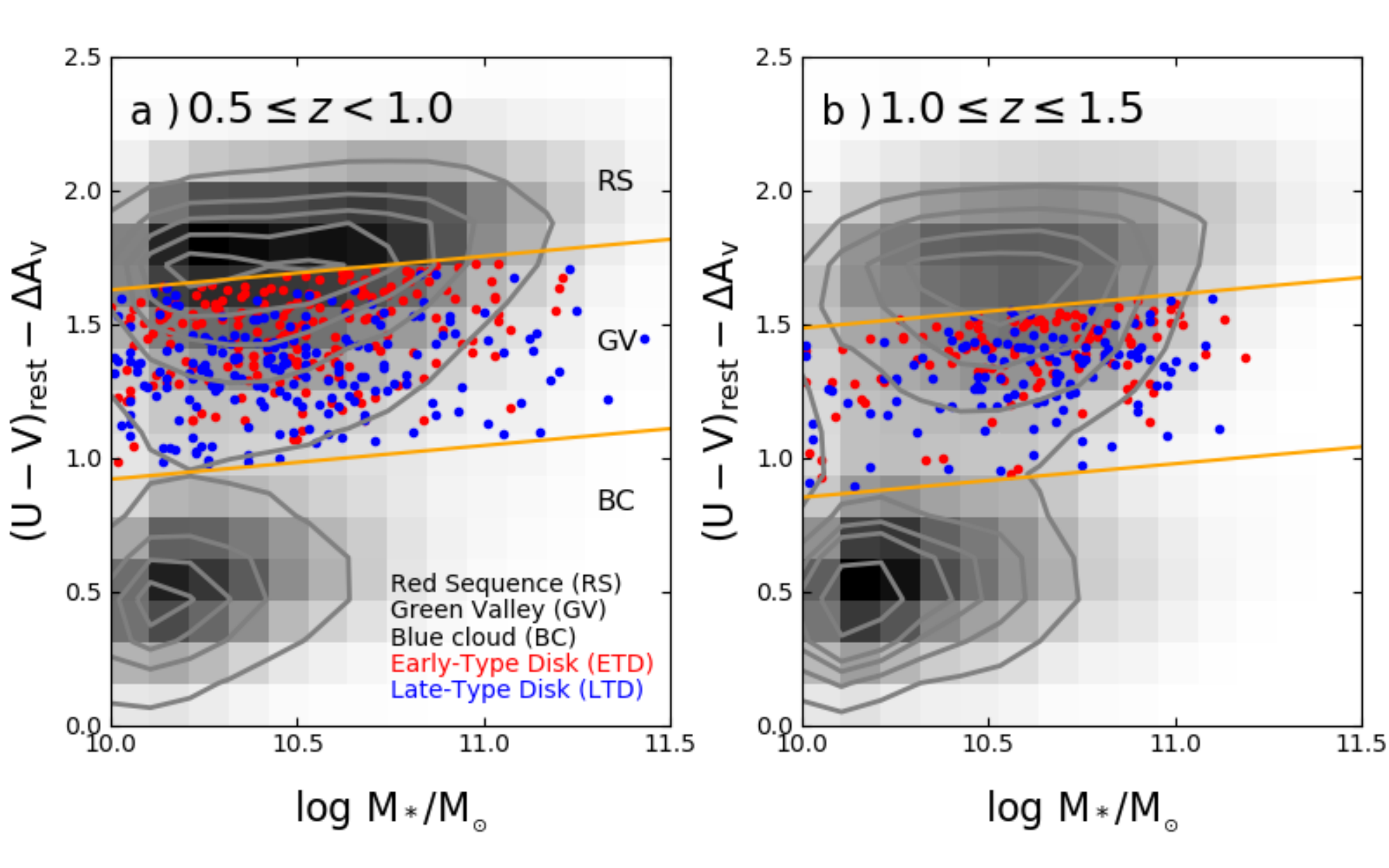}}
\caption{\label{fig01} Extinction-corrected rest-frame $U$-$V$ color as a function of stellar mass at $0.5 \leqslant z<1.0$ and $1.0 \leqslant z \leqslant 1.5$. The gray contours and blocks represent the relative density in the color-mass relation. The orange lines in each redshift bin represent the criteria of \cite{WangT+17} to separate galaxies into RS, GV, and RS. In the sample of GV galaxies, we utilize the morphological classification from \cite{Huertas-Company+15b} to divide the GV galaxies into ETD (red circles) and LTD (blue circles) galaxies after removing AGN candidates and building a``mass-matched'' sample (see Sections \ref{sec2.2:morphology} and \ref{sec2.3:sample selection}).}
\end{figure*}

For constructing the sample of GV galaxies precisely, we adopt the color-based separation criteria from \cite{WangT+17}. Inspired by \cite{Brammer+09}, \cite{WangT+17} consider the extinction correction to avoid the blend of quiescent RS galaxies and dusty SFGs. The criteria for separating galaxies into BC, GV, and RS galaxy populations are as follows:
\begin{eqnarray} \nonumber
 (U - V )_{\rm rest} - \Delta A_V = 0.126 \log (M_*/M_{\sun}) + 0.58 - 0.286z; \,\, \nonumber  \\
\nonumber
 (U - V )_{\rm rest} - \Delta A_V = 0.126 \log (M_*/M_{\sun})  - 0.24 - 0.136z, \,\, \nonumber
\end{eqnarray}
where $\Delta A_V=0.47 A_V$ is the extinction correction of rest-frame $U - V$ color, and the correction factor  0.47 is determined by the \cite{Calzetti+00} extinction law. The GV galaxies are distributed in the middle of the double lines.  In this work, we first select 1316 GV galaxies at $0.5 \leqslant z \leqslant 1.5$, and utilize the morphological classification of galaxies in \cite{Huertas-Company+15b} to select 376 ETD and 436 LTD galaxies (see section \ref{sec2.2:morphology} for details).

Considering the AGN effect on galaxy morphology and avoiding the complexity of GV evolutionary mechanisms due to AGN feedback, we eliminate AGN candidates from the sample of GV galaxies as much as possible. The X-ray sources are classified as AGN candidates by \cite{Xue+16} in Chandra Deep Field-North (CDF-N) and \cite{Luo+17} in Chandra Deep Field-South (CDF-S), which satisfy at least of one of the following criteria: (1) the intrinsic X-ray luminosity $L_{\rm 0.5-7kev}\geqslant 3\times 10^{42}$ erg s$^{-1}$; (2) the effective photon index $\Gamma_{\rm eff} \leqslant 1.0$; (3) the X-ray-to-optical flux ratio $\log(f_{\rm X}/f_{\rm R})>-1$, where $f_{\rm X}$=$f_{\rm 0.5-2kev}$, $f_{\rm 2-7kev}$, or $f_{\rm 0.5-7kev}$, and $f_{\rm R}$ represents the observed-frame R-band flux; (4) $L_{\rm 0.5-7kev}/L_{\rm 1.4 GHz}\geqslant2.4 \times 10^{18}$, where $L_{\rm 1.4 GHz}$ represents the rest-frame 1.4 GHz monochromatic luminosity in units of W Hz$^{-1}$; (5) identified as AGN by spectral certification (more detail to see \citealt{Xue+11}).
By cross-matching the catalogs from \cite{Xue+16} and \cite{Luo+17} within a radius of $\rm 1.\arcsec 5$, we remove those X-ray AGN candidates from our GV sample. Besides, the mid-infrared (MIR) color criterion as a potentially powerfully technique is gradually accepted by many studies to identify obscured AGN candidates (\citealt{Fang+15,Bornancini+17,Chang+17,Lu+20}), because the MIR emission is relatively insensitive to the intervening obscuration, so that it can trace the reprocessed radiation of dust heated by its central nucleus (\citealt{Lacy+04,Lacy+07,Stern+05,Donley+12}). Therefore, we also adopt the MIR color criteria from \cite{Donley+12} to get rid of MIR AGN candidates. After removing  a total of 66 AGN candidates, there are 216 ETD and 200 LTD galaxies left at $0.5\leqslant z < 1.0$, and 119 ETD and 211 LTD galaxies left at $1.0\leqslant z \leqslant 1.5$. It is worth noting that within observation limits, although we have adopted as much as possible criteria to eliminate the effect of AGN candidates on our sample, it is still hard to remove some sources with weak AGN activity or star-forming feedback. Thus, these potential sources can also fit well with the following results.

Subsequently, in order to minimize the stellar mass effects on the statistics of the morphological analysis, we further build a ``mass-matched'' sample of massive galaxies at $0.5\leqslant z\leqslant 1.5$ to ensure that the selected ETD and LTD galaxies have the same distributions of redshift and stellar mass. Following \cite{Gu+19}, we also define the minimum count among two populations as the number of galaxies in the mass-matched sample in fixed redshift and stellar mass bins. To be specific, for every galaxy with ($z_0$, $M_{*,0}$) in the ETD galaxies with the minimum count, the unique one with the smallest value of $\sqrt{(z_i-z_0)^2+(\log M_{*,i}-\log M_{*,0})^2}$ is selected from the other LTD galaxies into the mass-matched sample. Finally, our sample consists of 200 ETD and 200 LTD galaxies at $0.5\leqslant z < 1.0$, and  119 ETD and 119 LTD galaxies at $1.0\leqslant z \leqslant 1.5$.
As shown in Figure \ref{fig01}, the extinction-corrected rest-frame color as a function of stellar mass is present. The orange lines represent the separation criteria from \cite{WangT+17}. The region between two orange lines is called as the GV, and the upper and lower regions are the RS and BC, respectively. The ETD (red) and LTD (blue) galaxies in our GV sample are evenly distributed in the middle region.

\section{Analysis}
\label{sec3:analysis}
In order to understand the impact of the MQ mechanism on the star formation activity of GV galaxies, the comparison between ETD and LTD galaxies in the concentration, sSFR, morphology quenching efficiency $Q_{\rm mor}$, and environment will be analyzed and discussed.
\subsection{Concentration}
\label{sec3.1:Conc}
Nonparametric measurements are performed on the NIR images to derive the concentration index ($C$) by using the \textit{Morpheus} software (\citealt{Abraham+07}). The concentration index $C$, defined by \cite{Abraham+1994}, can be calculated by the ratio between the integral flux within the inner isophotal radius, 0.3\textit{R}, and the integral flux within the outer isophotal radius, \textit{R}. For a given galaxy with a high $C$, this implies that the light of a galaxy is concentrated at the center, and a galaxy with a higher $C$ might further possess a centrally concentrated bulge, which can bind more masses through its central gravitational potential.

\begin{figure*}[htb]
\center{\includegraphics[width=15cm] {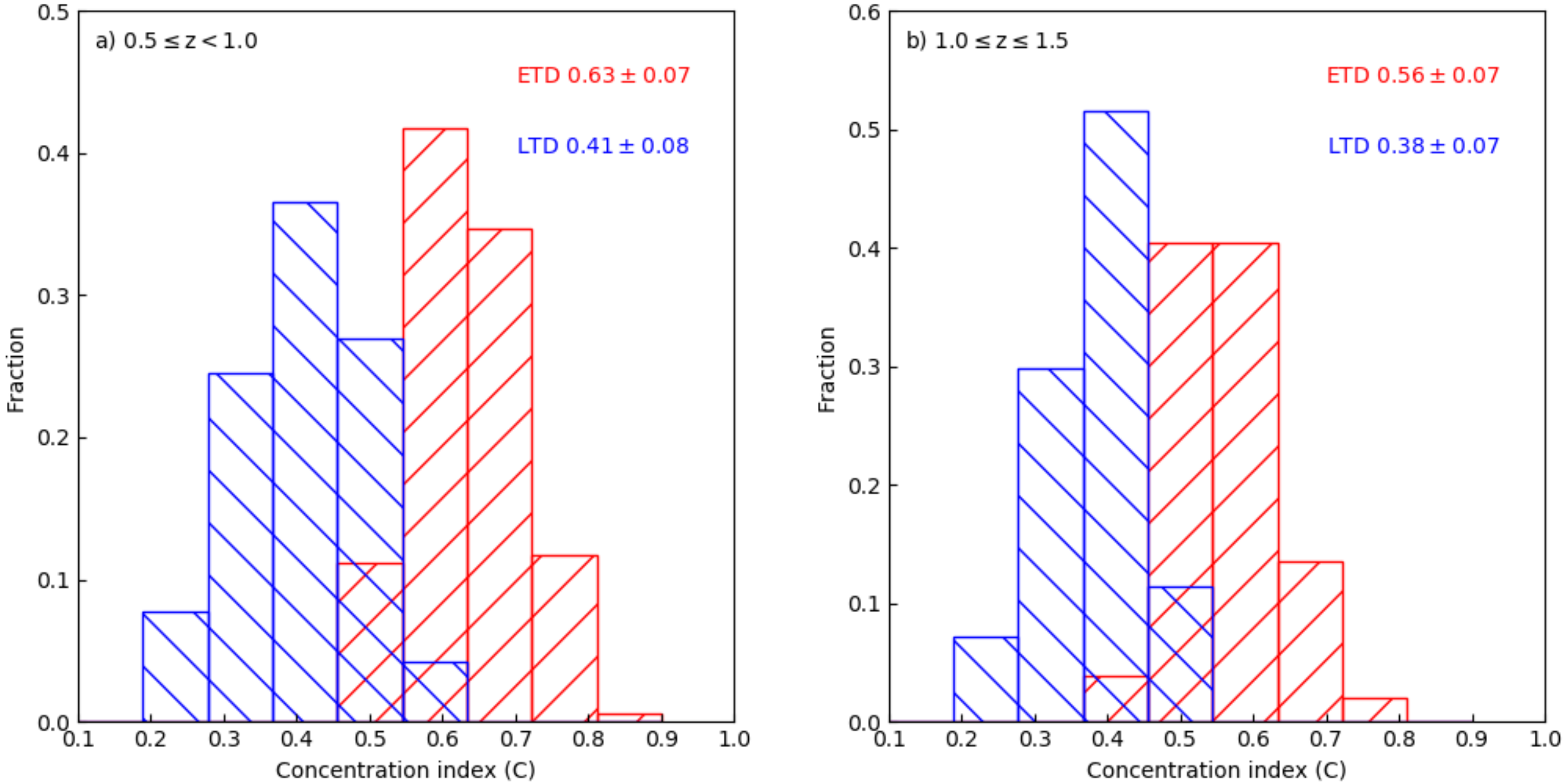}}
\caption{\label{fig02} Hatched histograms of the concentration index of ETD (red) and LTD (blue) galaxies in two redshift bins. The mean value and standard deviation are given in the upper right corner of each panel. }
\end{figure*}

The histograms of the concentration index of ETD (red) and LTD (blue) galaxies in GV at $0.5\leqslant z<1.0$ and $1.0 \leqslant z \leqslant 1.5$ are shown in Figure \ref{fig02}. This presents that the histograms of ETD galaxies are obviously distributed at the higher end of concentration index $C$, whereas the histograms of LTD galaxies place at the lower end. As expected, ETD galaxies possess a higher concentration, which is in line with the result of \cite{van+der+Wel+14}, who find that ETGs are on average smaller than LTGs. The corresponding mean values and standard deviations of ETD and LTD galaxies are also present in the upper right corner of the figure. From the higher- to lower-redshift bin, it can be found that both the mean values of the concentration indices of ETD and LTD galaxies become larger and the concentration difference between them seems to be larger. Similar results also can be found within a fixed stellar mass bin (i.e., $10.0 \leqslant \log(M_*/M_{\odot})<10.4$, $10.4 \leqslant \log(M_*/M_{\odot})<10.8$, and $\log(M_*/M_{\odot})\geqslant10.8$). As mentioned in  panels (e)-(h) of Figure 5 from \cite{Gu+18}, the morphological transformation of GV galaxies with cosmic time is accompanied by the growth of the bulge component as well.

\subsection{Specific Star Formation Rate}
\label{sec3.2:sSFR}
Trends between the sSFR and morphology have been studied in many previous works in which the sSFR of a galaxy is anticorrelated with central mass concentration or the presence of a bulge (e.g., \citealt{Kauffmann+03b,Franx+08,Wuyts+11b,Lang+14,Whitaker+15,McPartland+19}). For instance, earlier-type spiral galaxies have lower sSFRs than later-type spirals (\citealt{Parkash+18}). The lower sSFRs of massive galaxies may indicate a lower star formation efficiency rather than a lower gas fraction (\citealt{Schreiber+16}). Therefore, we compare the sSFR distribution of ETD galaxies with that of LTD galaxies to figure out whether the morphological suppression due to the presence of a central bulge has an influence on the sSFR.

\begin{figure*}[htb]
\center{\includegraphics[width=15cm] {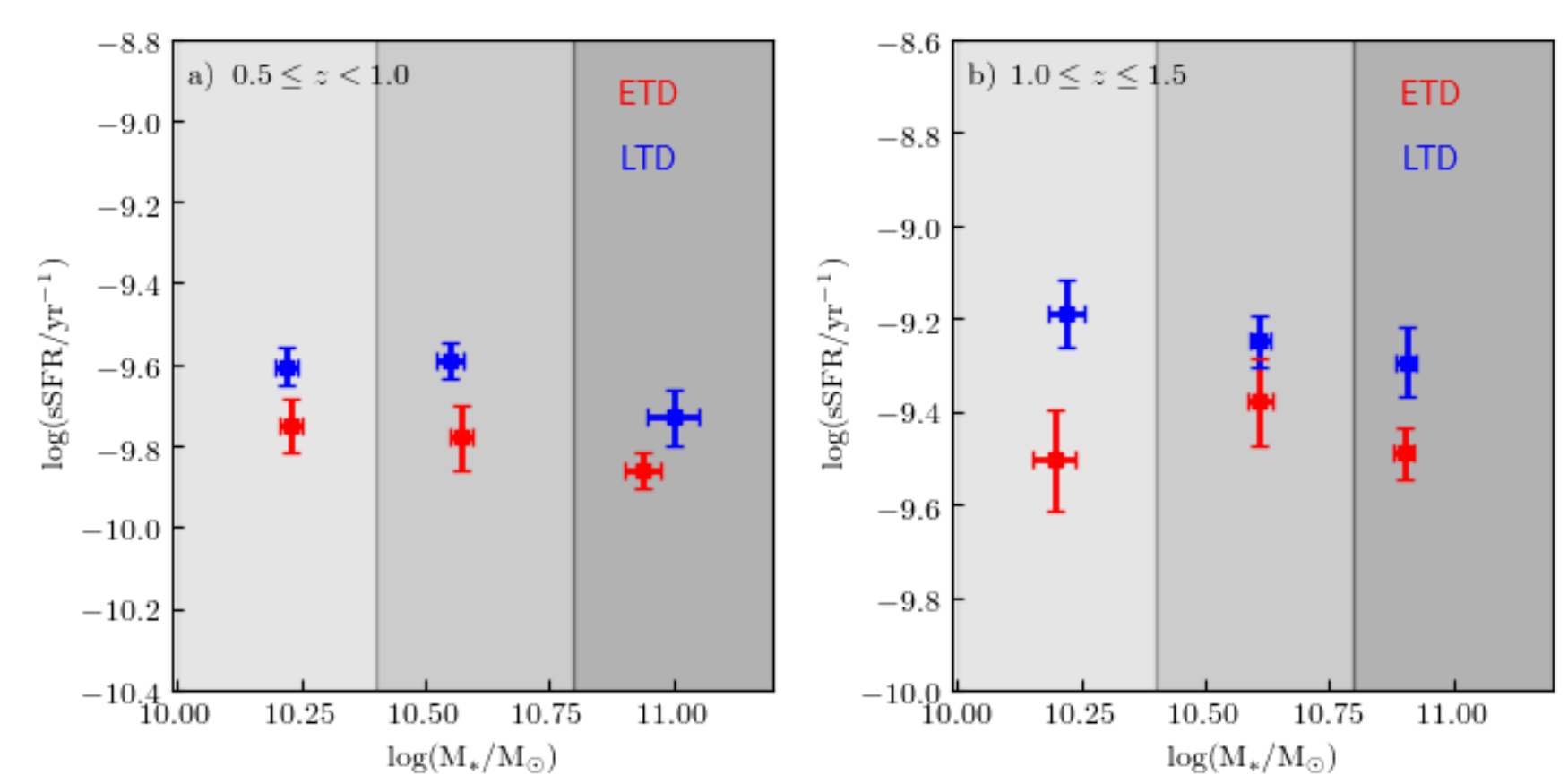}}
\caption{\label{fig03} The sSFR-stellar mass relation for ETD (red) and LTD (blue) galaxies at $0.5 \leqslant z <1.0$ and $1.0 \leqslant z \leqslant 1.5$. In each redshift bin, every class of galaxy is separated into three stellar mass bins (i.e., $10.0 \leqslant \log(M_*/M_{\odot})<10.4$, $10.4 \leqslant \log(M_*/M_{\odot})<10.8$, and $\log(M_*/M_{\odot})\geqslant10.8$). The data point represents the median values of the distribution for ETD and LTD galaxies, respectively. The standard errors of medians are estimated by bootstrapping the samples in each stellar mass bin 1000 times.}
\end{figure*}

Figure \ref{fig03} shows the sSFR-stellar mass relation at $0.5 \leqslant z \leqslant 1.5$, considered in three stellar mass bins. Comparing the relation of ETD (red) galaxies with that of LTD (blue) galaxies, it is apparent that ETD galaxies possess lower sSFR than LTD galaxies regardless of the different stellar mass bins, which is consistent with previous works (e.g., \citealt{Kauffmann+03b,Franx+08,Wuyts+11b}). It also might hint that the presence of a central bulge is conducive to suppress the star formation activity of a galaxy. With the increase of the stellar mass, both ETD and LTD galaxies have slightly decreasing signs of their star formation, especially for massive ETD galaxies. This implies that massive bulge-dominated GV galaxies may spend a shorter time quenching their star formation than less massive GV galaxies, which is also in line with the ``downsizing'' scenario, where massive SFGs are formed and subsequently quenched at an earlier epoch than those less massive SFGs (e.g., \citealt{Cowie+1996,Noeske+07,Goncalves+12,Gu+18}).
Besides, if we ignore the red point with a larger error bar for the ETD galaxies in the lowest stellar mass bin at $1.0 \leq z \leq 1.5$, it can be found that both the sSFR of ETD galaxies and that of LTD galaxies are getting smaller with cosmic time, which might suggest that the GV galaxies are gradually quenching their star formation into the red sequence with the gradual consumption of gas reserves.

\subsection{Morphology Quenching Efficiency}
\label{sec3.3:Qmor}
It has long been known that SFGs evolve into the passive galaxies by quenching their star formation, which is usually accompanying morphological transformation (e.g., \citealt{Kauffmann+03b,Wuyts+11b,Whitaker+15}).
From the studies of the S\'{e}rsic index and stellar mass profiles, it has been found that GV galaxies host larger bulges than BC galaxies at the same stellar mass (\citealt{Schiminovich+07,Fang+13}).
The so-called MQ scenario thus has been proposed for the star formation quenching mechanism (\citealt{Martig+09}). The description of MQ has always been qualitative for a long time,  whereas it lacks a quantitative description and discussion instead. Therefore, in order to quantify the contribution of the galaxy bulge to the suppression of star formation, we try to define a simple morphology quenching efficiency ($Q_{\rm mor}$), a dimensionless parameter, as the deficiency in the ratio of the median sSFR of ETD galaxies and LTD galaxies compared with the unity:\\
\begin{equation}\label{eq3:Qmor}
  Q_{\rm mor} = \rm 1 - \frac{sSFR(ETD)}{sSFR(LTD)}.
\end{equation}
Removing the effect of AGNs on our sample as much as possible, if the presence of the bulge of a galaxy can help quench its star formation, it will result in that the median sSFR of ETD galaxies is lower than that of LTD galaxies. Subsequently, the MQ can be reflected by the increase of $Q_{\rm mor}$. When the value of $Q_{\rm mor}$ is close to zero, it implies that the morphology quenching may not be important for GV galaxies, as claimed by \cite{Koyama+19} for local GV galaxies.

\begin{deluxetable}{cccccc}
\tablecaption{Morphology Quenching Efficiency ($Q_{\rm mor}$) in Three Stellar Mass Bins.
\label{tab1:Qmor}}
\tablehead{\colhead{Redshift Range} & \colhead{Stellar Mass Range} & \colhead{$Q_{\rm mor}$}}
\startdata
$0.5\leqslant z < 1.0$         &$10.0\leqslant \log(M_* / M_{\sun}) < 10.4$&$0.29\pm0.07$ \\
$                    $         &$10.4\leqslant \log(M_* / M_{\sun}) < 10.8$&$0.34\pm0.08$ \\
$                    $         &$\log(M_* / M_{\sun}) \geqslant 10.8$&$0.26\pm0.06$ \\
$1.0\leqslant z\leqslant 1.5$  &$10.0\leqslant \log(M_* / M_{\sun}) < 10.4$&$0.52\pm0.12$ \\
$                           $  &$10.4\leqslant \log(M_* / M_{\sun}) < 10.8$&$0.26\pm0.05$ \\
$                           $  &$\log(M_* / M_{\sun}) \geqslant 10.8$&$0.41\pm0.08$ \\
\enddata
\end{deluxetable}

\begin{figure*}[htb]
\center{\includegraphics[width=15cm]{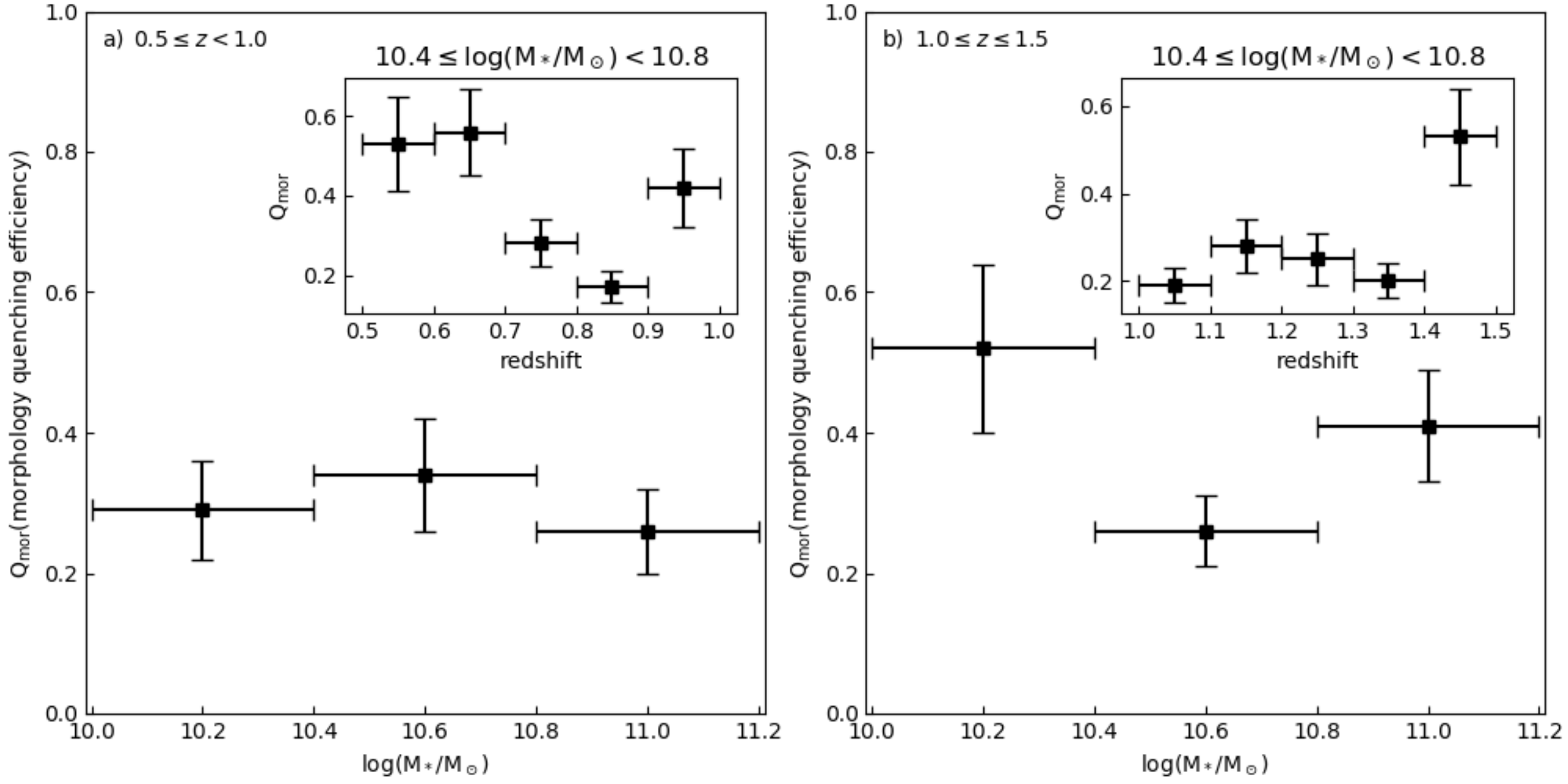}}
\caption{\label{fig04} Distributions of the morphology quenching efficiency $Q_{\rm mor}$ in three different mass bins at $0.5 \leqslant z <1.0$ and $1.0 \leqslant z \leqslant 1.5$ (see Table \ref{tab1:Qmor}). The redshift evolution of $Q_{\rm mor}$ in the moderate mass bin (see Table \ref{tab1:Qmor}) is shown in the inset panels. The corresponding error is derived by using the error propagation, assuming that  the standard deviation of sSFR in each stellar mass bin represents its corresponding error of the sSFR median.}
\end{figure*}

In two redshift intervals, the $Q_{\rm mor}$ is calculated in three stellar mass bins and the corresponding values are given in Table \ref{tab1:Qmor}. Figure \ref{fig04} also presents the distribution of $Q_{\rm mor}$ in different stellar mass bins. From both Figure \ref{fig04} and Table \ref{tab1:Qmor}, it can be found that the $Q_{\rm mor}$ is always larger than 0.2 regardless of the difference in stellar mass.
Considering the dependence of $Q_{\rm mor}$ on redshift, we recalculate $Q_{\rm mor}$ in smaller different redshift bins and find that the morphology quenching efficiency $Q_{\rm mor}$ is not sensitive to redshift.
The inset panels of Figure \ref{fig04} show for a fixed redshift bin of $\Delta z$=0.1 the redshift evolution of $Q_{\rm mor}$ in the typical stellar mass bin ($10.4 \leqslant \log{M_*/M_\odot} < 10.8$).
On average, all $Q_{\rm  mor}$ are greater than 0.2 within the error range. The above results may suggest that within the redshift range from 0.5 to 1.5 the morphology quenching efficiency $Q_{\rm mor}$ seems to be insensitive to the different stellar mass and redshift.

\begin{figure*}[b]
\center{\includegraphics[width=13cm] {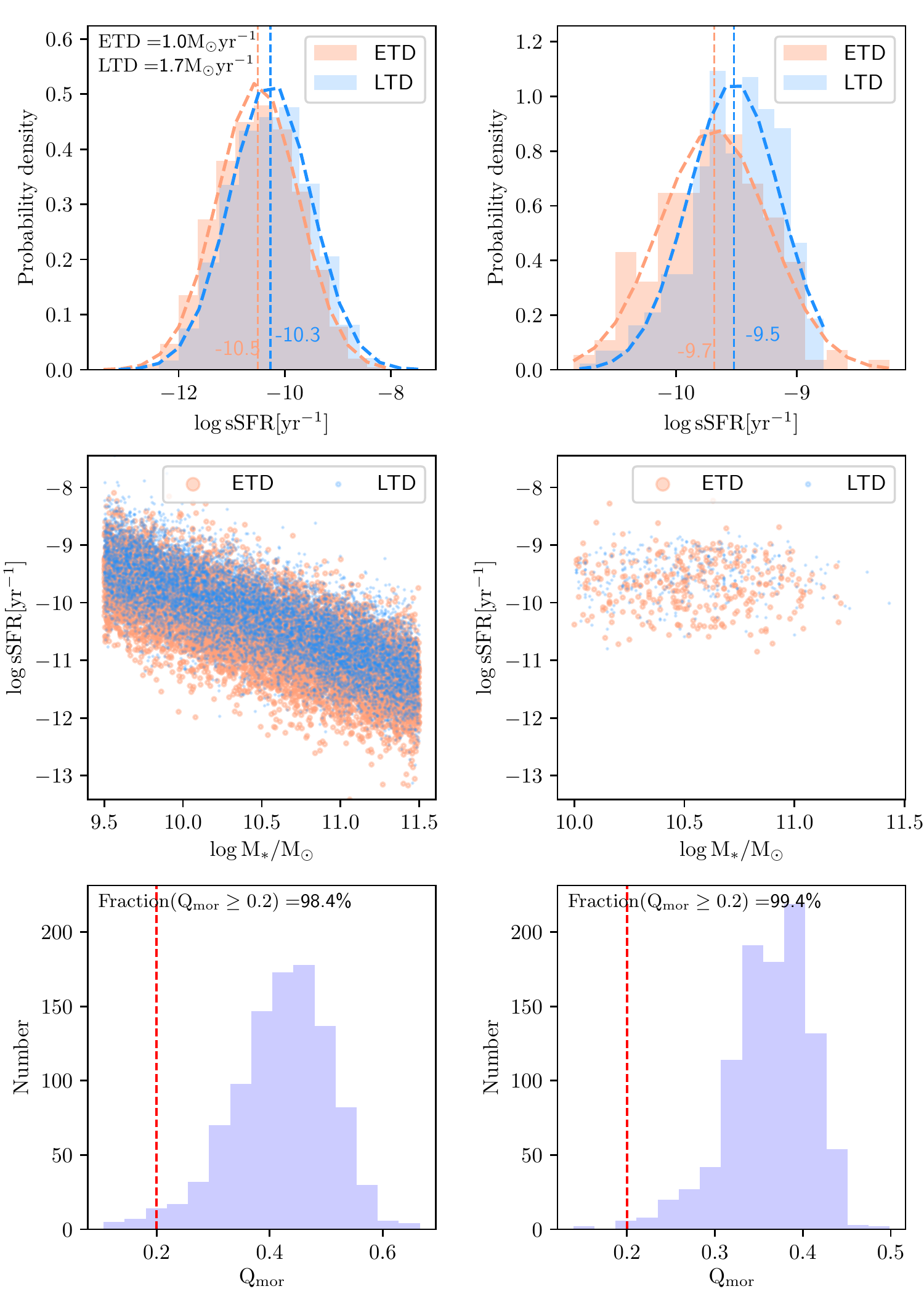}}
\caption{\label{fig05} Left panels: assuming normal distributions of SFR with stellar mass spanning uniformly from $10^{9.5}$ to $10^{11.5} M_\odot$, the sample for each type consists of 10,000 galaxies with different SFR mean, shown in the upper left corner. The normalized histograms of sSFR for ETD (salmon) and LTD (blue) galaxies are fitted by normal distributions (dotted curves), shown in the top panel. The corresponding mean values of sSFR are shown near the vertical dotted lines. The middle panel shows the scatter of stellar mass as a function of sSFR. The bottom panel shows the histogram of $Q_{\rm mor}$, which is obtained by randomly taking 500 samples from each type of galaxy 1000 times. The red dotted lines present $Q_{\rm mor}=0.2$, and the fraction of $Q_{\rm mor} \geq 0.2$ is shown at the top of its panel. Right panels: similar to the left panels, but for the real sample in this work, consisting of 319 ETD and 319 LTD galaxies.}
\end{figure*}

Besides, in order to measure the effect of SFR on the value of $Q_{\rm mor}$, a quantitative experiment is adopted. First, each type of mock sample consists of 10,000 galaxies with stellar masses spanning uniformly from $10^{9.5}$ to $10^{11.5} M_\odot$. And normal distributions of $\rm \log (SFR)$ for both are assumed with different means but the same variances $\sigma \sim 0.5$, which are derived by fitting our real ETD and LTD subsamples. As shown in Figure \ref{fig05}, the top panels present the sSFR's distributions of the ETD (salmon) and LTD (blue) galaxies for the mock (left) and real (right) sample, respectively, and the corresponding scatter diagrams between sSFR and stellar mass are shown below. Then, we randomly choose 500 galaxies from each type of sample at a time to calculate the corresponding median value of sSFR and $Q_{\rm mor}$. After repeating this process 1000 times, as shown in the bottom of Figure \ref{fig05}, when the difference of the mean SFR between ETD and LTD galaxies is about $\Delta_{\rm SFR}=0.7  M_\odot \rm yr^{-1}$, about 98.4$\%$ of $Q_{\rm mor}$ are greater than 0.2, which is comparative to that fraction (99.4\%) by repeating the process for our real sample. And the fraction tends to be smaller ($91.5\%$) with the increasing variance $\sigma \sim 1.0$. While both the intrinsic scatter and uncertainties of SFR included in its variance, are hardly separated clearly, the more precise criterion of $Q_{\rm mor}$ thus is not easily obtained. Therefore,
the degree of morphological transition on the quenching of GV galaxies at $0.5\leqslant z\leqslant1.5$ in this work tends to be described by the value of the morphology quenching efficiency $Q_{\rm mor}$ of 0.2 with the difference of the mean SFR between ETD and LTD galaxies $\Delta_{\rm SFR}=0.7  M_\odot \rm \; yr^{-1}$, which is a quantitative value compared to before.

\subsection{Environment}
\label{sec3.4:env}
The observed properties of galaxies are generally considered to be related to their host environment. In the local universe, it has been confirmed that early-type passive galaxies often live in dense environments (e.g., groups and clusters), whereas in less dense environments (i.e., field) most galaxies are late-type and star-forming (e.g., \citealt{Dressler+1980,Balogh+04,Kauffmann+04,Peng+10,Woo+13,Abraham+07}).
While this situation remains controversial at intermediate to high redshifts, the star formation of galaxies is found to be enhanced in a dense environment (e.g., \citealt{Elbaz+07,Cooper+08}), or to be not significant related to the environment (e.g., \citealt{Darvish+16}), or to be similar to the relation as in the local universe (e.g., \citealt{Patel+09}). Compared to the ``MQ'' (an internal process), the ``environmental quenching'' (an external process) is also a relevant mechanism to reduction and cessation of star formation in a galaxy. In order to further discern which mechanism is the most dominant in quenching galaxies, we improve the traditional measurement of the environment (as mentioned in Gu et al. 2020) and compare the environment around ETD galaxies to that around LTD galaxies.

\begin{figure*}[htb]
\center{\includegraphics[width=13cm] {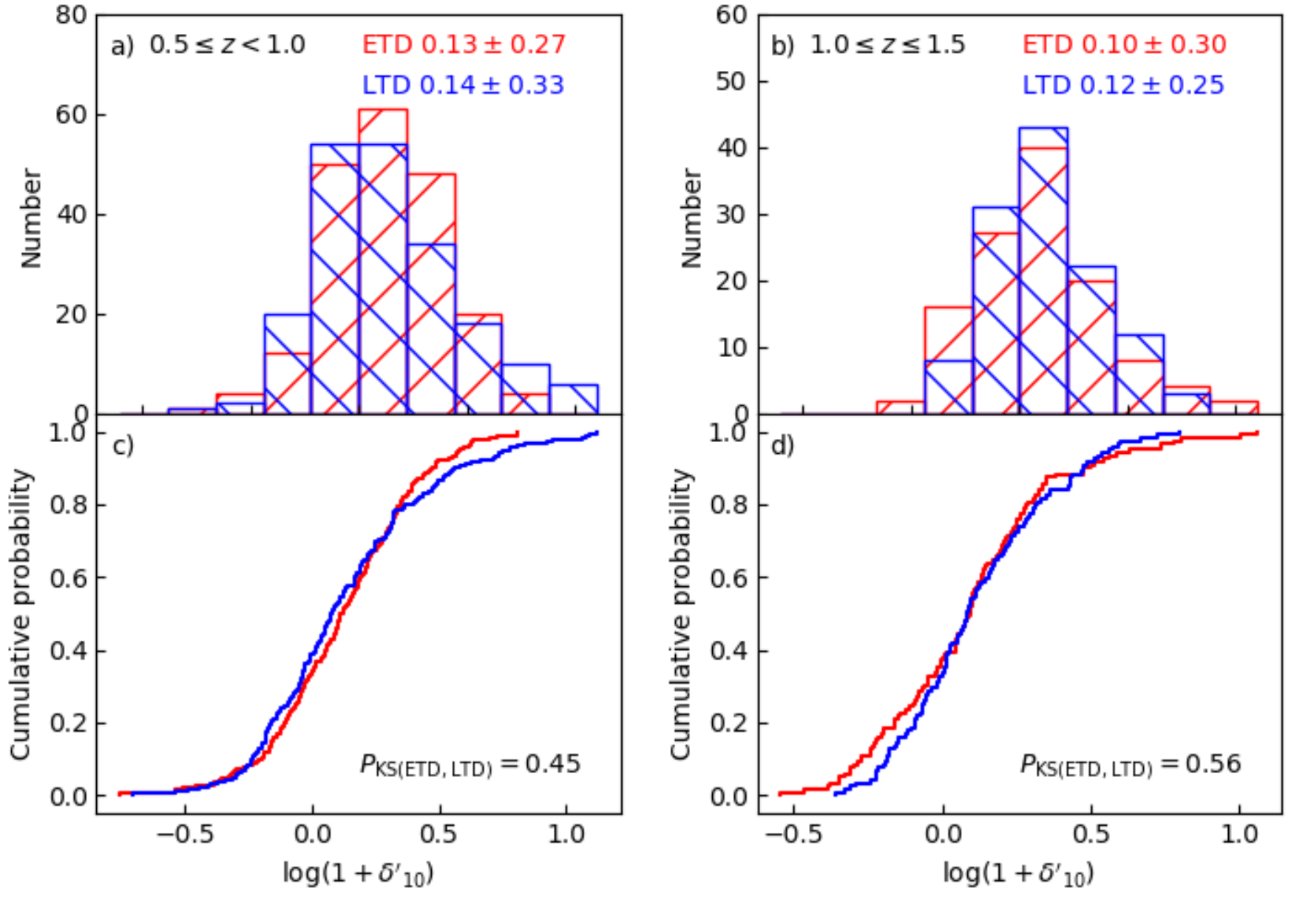}}
\caption{\label{fig06} Distributions (top panels) and cumulative probabilities (bottom panels) of overdensity in two redshift bins. The ETD and LTD galaxies are denoted by red and blue hatched histograms, respectively. Their corresponding mean values and standard deviation are given in the top-right panels. The probabilities of K-S tests between two subsamples are marked in the bottom-right panels.}
\end{figure*}

Instead of the traditional indicator of local environment, which often relies on the 10th nearest neighbor or the total number of near galaxies in a fixed radius (\citealt{Dressler+1980}), we have modified the environmental indicator using the Bayesian metric to consider the distances of all \textit{N} neighboring galaxies (\citealt{Ivezic+05,Cowan+Ivezic+08}). The method can improve the accuracy of density estimation by 3.5 times over the reconstructed probability density distribution (see \citealt{Ivezic+05} for details).
To ensure the high accuracy of redshift in the 3D-HST/CANDELS (\citealt{Skelton+14}),  a magnitude-limited ($H_{\rm F160W}$$<$$25$ ) sample at $0.5 \leqslant z \leqslant 1.5$ is constructed first, and then we can estimate the local surface density of every galaxy in the sample by $\Sigma'_{10}=1/(\Sigma^{10}_{i=1}d^2_{i})$, where $d_{i}$ is the projected distance to the \textit{i}th nearest neighbor within a redshift slice ($|\Delta z|<\sigma_{z}(1+z)$, $\sigma_{z}=0.02$). Subsequently, we define the dimensionless overdensity, $1+\delta'_{10}$, as the environment indicator:
\begin{equation}\label{eq4:env}
1+\delta'_{10} = \frac{\Sigma'_{10}}{\langle \Sigma'_{10}\rangle}=\frac{\Sigma'_{10}}{k'_{10}\Sigma_{\rm surface}},
\end{equation}
where $\Sigma_{\rm surface}$/arcmin$^{-2}$ is the surface number density within a given redshift slice. The denominator $\langle \Sigma'_{10}\rangle$, a typical tracer of environment, can be evaluated by multiplying the correction factor $k'_{10}$ and the local surface density $\Sigma'_{10}$, where the adopted value of $k'_{10}$ is equal to 0.06. And the determination of overdensity has been proven to be competent to trace the high-redshift structure by reseeking some known galaxy groups and cluster candidates, such as J021734-0513 at $z \sim 0.65$ (\citealt{Galametz+18}) and CIG 0218.3-0510 at $z\sim 1.62$ (\citealt{Papovich+10}). 

The overdensity distributions of ETD (red) and LTD (blue) galaxies in two redshift bins are shown in the top panels of Figure \ref{fig06}. Both ETD and LTD galaxies have similar histograms and mean values of the environmental tracer. In order to illuminate the difference between their environments, the corresponding cumulative distribution functions are given in the bottom panels. And the  Kolmogorov-Smirnov (K-S) tests between ETD and LTD galaxies in two \emph{z}-bins are performed to see whether both subsamples are drawn from the same underlying parent distribution. The upper limit of the probability of K-S test, quantitative$P_{\rm KS(ETD, LTD)}$, is 0.05, verifying that two samples have different distributions of environment at $\geqslant 2\sigma$ significance. In Figure \ref{fig06}, the high probabilities of the K-S test between ETD and LTD galaxies reveal that there is no difference between the environments surrounding the ETD and LTD galaxy, which are still established within fixed stellar mass bins.
Compared to LTD galaxies, ETD galaxies tend to have lower sSFR, but both reside in a similar environment, which might suggest that the quenching of star formation for massive ($M_* > 10^{10} M_{\odot}$) GV galaxies at $0.5 < z <1.5$ is not predominated by the environment. Based on the same CANDELS, \cite{Ge+20} and \cite{Chartab+20} both have confirmed that the environmental quenching is more efficient for low-mass ($M_* < 10^{10} M_{\odot}$) galaxies at $z \lesssim 1.5$, rather than for massive ($M_* > 10^{10} M_{\odot}$) galaxies, which agrees with our result.

\section{Discussion}
\label{sec4:discussion}
Much of the work on GV galaxies has found that GV galaxies are expected to be in the star formation quenching phase (\citealt{Bell+04,Faber+07,Martin+07,Gu+18,Gu+19}), which is suitable for studying the impact of galaxy morphologies on the quenching of star formation.
Thus, we construct two subsamples of ETD and LTD galaxies in the GV stage at the intermediate redshifts to compare their concentrations, sSFRs, morphology quenching efficiencies $Q_{\rm mor}$, and environment. Actually, both internal and external processes are closely related to the suppression of the star formation of a galaxy. The internal processes often refer to those that are relevant to individual galaxy properties, such as the morphology and the presence of an AGN, while in this work, regardless of the effect of some sources with weak AGN activity or star-forming feedback, we have eliminated as much as possible the influence of AGNs on host galaxies (see section \ref{sec2.3:sample selection} for details), just leaving the impact of the morphology of the galaxy.
In contrast, the external processes are usually relative to the surrounding environment of a galaxy.
In section \ref{sec3:analysis}, the comparisons of ETD and LTD galaxies have suggested that the environments surrounding both ETD and LTD galaxies are similar, whereas ETD galaxies possess higher concentration and lower sSFR than LTD galaxies.
This might hint that it is the impact of the galaxy morphology, not the environment, that may play a more important role in the star formation quenching of the massive ($M_* \geqslant 10^{10} M_\odot$) GV galaxies at $0.5 \leq z \leq 1.5$, which is consistent with the results of \cite{Gu+19} and \cite{Ge+20}. And it also agrees with the previous result that low-mass ($M_* <10^{10} M_\odot$) galaxies at $z<1.5$ is more dependent on the environmental effect (\citealt{Scoville+13,Lee+15,Guo+17}). Note that those comparisons between ETD and LTD galaxies are under similar stellar mass and redshift distributions (see section \ref{sec2.3:sample selection} for details).

To quantify the effect of quenching on star formation in a galaxy, we define a parameter of the morphology quenching efficiency $Q_{\rm mor}$ to correlate the sSFR of the galaxy with its morphological transformation.
In Figure \ref{fig04},  it is shown that the $Q_{\rm mor}$ has no apparent dependence on stellar mass and redshift, and the value of $Q_{\rm mor}$ is always larger than 0.2 within the error range. Meanwhile, a quantitative experiment on the SFR's effect on the measured $Q_{\rm mor}$ indicates that, assuming a uniform stellar mass distribution and a normal distribution of SFR for each type, the fraction of  $Q_{\rm mor} \geqslant 0.2$ can be as high as 98.4$\%$ when the difference between the mean SFRs of ETD and LTD galaxies is about $0.7 M_\odot \rm yr^{-1}$. It is comparable to that fraction ($99.4\%$) from our real GV sample, which might hint that the level of the difference between the mean SFRs of ETD and LTD galaxies in our GV sample tends to be about $0.7 M_\odot \rm yr^{-1}$, and
the level of the morphology quenching in our GV galaxies tends to be quantified by this value (i.e., $Q_{\rm mor}$=0.2). Generally, our results are in line with the morphological transformation sequence of bulge buildup for the similar GV galaxies from \cite{Gu+19}, which is seen as their star formation activities are gradually shut down at $0.5 <z<1.5$.
In contrast, \cite{Koyama+19} use 35 GV galaxies at $0.025 < z< 0.05$ to reveal their distributions of molecular gas mass and fraction and the star formation efficiency, while their results suggest that the presence of the stellar bulge does not reduce the efficiency of ongoing star formation in GV galaxies. The cause may be that their GV galaxies are selected from a small window in the $M_*$-sSFR plane (i.e., $10.8<\log(M_*/M_\odot) <11.2$ and $-11< \log(\rm sSFR/yr^{-1}) <-10.5$), in which the intrinsic effect of a galaxy bulge may not be well revealed. Compared to our higher stellar mass bin (i.e., $\log (M_*/M_\odot)\geqslant 10.8$), our median values of sSFR for ETD and LTD galaxies are larger than the upper limit of sSFR in \cite{Koyama+19}, and the $Q_{\rm mor}$ calculated by us are still more than 0.2.

Besides, it is worth mentioning that the quenching time of the GV sample at $0.5\leqslant z\leqslant1.5$ is longer than that at higher redshift. Previous works have found that the average galaxy quenching timescale decreases with increasing redshift (\citealt{Pandya+17,Gu+19}). Following the definition of the GV of \cite{WangT+17}, \cite{Lin+21} have introduced a cluster method to estimate the lifetimes of the RS, GV, and BC galaxies at $0.5\leqslant z\leqslant2.5$ by adopting six different halo mass functions (HMFs) from previous works (\citealt{Sheth+01,Reed+07,Tinker+08,Crocce+10,Watson+13} ). And it is found that the average lifetime of GV at $0.5<z<1.5$ is about 2 Gyr, while it is about 334 Myr at $1.5<z<2.5$ regardless of the adopted HMFs. Due to the same GV definitions used in this work, this perhaps implies that our GV galaxies go through a longer timescale of morphology quenching at $0.5\leqslant z\leqslant1.5$. Therefore, the results presented here tend to be biased toward GV galaxies with slower quenching rates.

 \section{Summary}
\label{sec5:summary}
In this work, we present an analysis of the effect of MQ on the massive ($M_* \geqslant 10^{10}M_\odot$) ETD and LTD green valley galaxies at $0.5\leqslant z \leqslant 1.5$ in all five 3D-HST/CANDELS fields by comparing their morphology, sSFR, and environment. Following are our main conclusions.

After eliminating the influence of AGNs as much as possible, we find that, regardless of whether it is within a fixed stellar mass bin, ETD galaxies have higher concentration index and lower sSFR, compared to LTD galaxies. This suggests that ETD galaxies possess the more concentrated morphologies and the more rapid star formation quenching.

Considering the external process to the galaxy star formation quenching, we also compare the environment surrounding the targeted ETD galaxies to that of LTD galaxies and find that there is no significant difference of environment between them. This hints that the environment may not be a dominated factor of quenching the star formation of massive ($M_*\geqslant 10^{10} M_{\odot}$) GV galaxies at $0.5\leqslant z\leqslant1.5$.

Excluding the effect of the external process of the environment, the MQ on the star formation then becomes more important qualitatively. To quantify it, we define a morphology quenching efficiency $Q_{\rm mor}$, a dimensionless parameter, to measure the correlation between morphology and star formation. Regardless of the stellar mass and redshift, when the difference between the mean SFRs of ETD and LTD galaxies is about 0.7 $M_\odot \rm yr^{-1}$, the fraction of $Q_{\rm mor}\gtrsim$0.2 is much higher than 90\%, which might suggest that the level of MQ in GV galaxies might be described by $Q_{\rm mor}\gtrsim$0.2.

\acknowledgments
This work is based on observations taken by the 3D-HST Treasury Program (GO 12177 and 12328) with the NASA/ESA HST, which is operated by the Association of Universities for Research in Astronomy, Inc., under NASA contract NAS5-26555. 
This work is supported by the Strategic Priority Research Program of Chinese Academy of Sciences?No. XDB 41000000), the National Key R\&D Program of China (2017YFA0402600), the National Natural Science Foundation of China (NSFC, Nos. 11973038, 11673004, 11873032, 1143305 and 11421303), the National Basic Research Program of China (973 Program, 2015CB857004) and the Chinese Space Station Telescope (CSST) Project.
G.W.F. acknowledges support from the Yunnan young and middle-aged academic and technical leaders reserve talent program (No.201905C160039), and Yunnan Applied Basic Research Projects (2019FB007). G.Y.Z. acknowledges support from China Postdoctoral Science Foundation funded project (2020M681281).
\bibliography{reference}{}
\bibliographystyle{aasjournal}
\end{document}